\documentclass[preprint,showpacs,preprintnumbers,amsmath,amssymb]{revtex4}%
\usepackage{graphicx}
\usepackage{dcolumn}
\usepackage{bm}
\usepackage{amsmath}
\usepackage{amsfonts}
\usepackage{amssymb}%
\usepackage{appendix}
\providecommand{\U}[1]{\protect\rule{.1in}{.1in}}
\newcommand{\be}{\begin{equation}}
\newcommand{\en}{\end{equation}}
\newcommand{\bea}{\begin{eqnarray}}
\newcommand{\ena}{\end{eqnarray}}
\begin{document}
\title{ Reconstructing k-essence: Unifying  the  attractor $n_S(N)$ and the
swampland criteria  }

\author{Ram\'on Herrera}
\email{ramon.herrera@pucv.cl} \affiliation{Instituto de
F\'{\i}sica, Pontificia Universidad Cat\'{o}lica de
Valpara\'{\i}so, Avenida Brasil 2950, Casilla 4059,
Valpara\'{\i}so, Chile.}

\date{\today}

\begin{abstract}

The reconstruction of a k-essence  inflationary universe,
 considering the unification between the swampland criteria
 and 
 the attractor given by 
the scalar spectral index   $n_S(N)$  together with  the slow roll parameter $\epsilon(N)$
 in terms of the number of $e$-folds  $N$
  is studied. In the context of   a coupling of the form 
$L(\phi)\,X$ in the k-essence model,  we find 
the effective potential $V$ and the
coupling parameter $L$ in terms of  
the scalar spectral index and the slow roll parameter under a 
general formalism.
  To apply the unification in our model, we consider  some examples 
 in order to rebuild the 
effective potential $V(\phi)$ and the coupling parameter
$L(\phi)$ as a function of the inflaton field $\phi$. Here, we find that the 
reconstruction gives rise to an exponential potential and also to natural and hyperbolic inflation, 
respectively. Thus, in this article 
 we show that it is possible to unify the theoretical foundations 
from the swampland criteria  
and the observational parameters corroborated by observations,  in 
the reconstruction of an inflationary universe.

\end{abstract}

\pacs{98.80.Cq}
\maketitle

\section{Introduction}

It is well known that the evolution of the early universe can be described by 
 the standard hot bing bang model, however this hot model  
presents some cosmological problems that  the inflationary stage or inflation  solves 
through  an accelerated  expansion previous to radiation era
 \cite{R1,R102,R103,Rin}. Nevertheless, the importance of  inflation is that this scenario  
 gives account of    the Large-Scale
Structure (LSS) \cite{R2,R203} and it  also  provides 
 a causal description
of the anisotropies observed in the Cosmic Microwave Background
(CMB) radiation \cite{astro,Hinshaw:2012aka,Ade:2013zuv,Planck2015,Ob2,DiValentino:2016foa}.

In the literature, we can find 
 different models that give account of   the
inflationary  evolution of the early universe. In this context, we  can distinguish 
the inflationary models where inflation is driven for  a canonical or non-canonical scalar field, see e.g.,
\cite{Ho,Kobayashi:2011nu,DeFelice:2013ar}.
 In this sense, 
we can stand out the k-essence inflationary model, where the description of k-essence  is  through
an action or Lagrangian density that  includes non standard higher order kinetic 
term associated to scalar field\cite{P1,P2}. An important consideration to take into account 
of the k-essence models is the fact that the  speed gravitational  waves 
  is equal to the speed of
light, coinciding with the speed obtained from 
 the detection of   gravitational waves by
GW170817 and  the $\gamma-$ray burst
\cite{TheLIGOScientific:2017qsa,Monitor:2017mdv,GBM:2017lvd}. Additionally, the k-essence models 
give  the possibility that the value of the speed of sound of the scalar perturbations 
is smaller than one or equal to one\cite{P1}. 
 In this form, the 
  k-essence model  is consistent with these observational data, 
  since the speed gravitational  waves 
  is equal to the speed of
light and  we can also have the possibility that the speed of sound associated to scalar perturbations 
could  be less 
than or equal to one, depending of the Lagrangian density associated to the 
k-essence model. Thus, 
 the   k-essence model can be interesting  to study the early (inflation) and current (dark energy)
 \cite{Melchiorri:2002ux}
 universe. In particular,   
  in the 
  context of inflation, different effective potentials associated to a scalar field 
   have been studied  under 
  the slow roll approximation \cite{Joyce:2014kja,DeSantiago:2012nk,Bose:2008ew}.

%

%%%%%%%%%%%%%%%%%%%%%%%%%%%%%%%%%%%%%%%%%%%%%%%%%%%%%%%%%%%%%%

% RECONSTRUCTION
%%%%%%%%%%%%%%%%%%%%%%%%%%%%%%%%%%%%%%%%%%%%%%

On the other hand,  the reconstruction of the
background   variables such as; effective potential, coupling functions, scale 
factor associated to the 
  inflationary models, from the 
observational parameters such as; the scalar spectrum,  scalar
spectral index  and the tensor to scalar ratio,   have
been analyzed   by several
authors\cite{H1,H2,H3,H4,M,Chiba:2015zpa,H5}. 
 In this sense,  a possible methodology for the 
reconstruction of 
inflation  under   
 the slow roll approximation, can be developed by means of  
 the parametrization of these cosmological parameters   
  or  attractors, in terms of the number of $e-$foldings $N$.

As  an example of  this methodology,  we have the
scalar spectral index  $n_S(N)$ as a function of the number $N$. In particular, the simple 
 parametrization or attractor $n_S(N)= 1-2/N$ 
 is well corroborated by Planck data \cite{Planck2018}, 
when the number of $e-$foldings $N\simeq$ 50 $\sim$ 60. Here we consider that the number of 
$e-$foldings $N\simeq$ 50 $\sim$ 60
 corresponds to the comoving scale
$k$ crossed the Hubble radius i.e., $k=aH$ during the inflationary epoch.

In the framework of  
  the General Relativity (GR), the reconstruction of inflation under this procedure gives origin to 
  different inflationary models according to the 
   attractor  point $n_S(N)$ given by   $n_S(N)= 1-2/N$, 
during the slow roll scenario for large $N$. In this way, 
 we can  have; the hyperbolic tangent model or 
T-model \cite{T}, 
 E-model\cite{E},
$R^2$-model\cite{R102}, the chaotic inflationary model\cite{R103}, the study of 
 Higgs inflation \cite{Higgs,Higgs2}, etc. In the reconstruction of two 
 background 
 variables as the case of
  warm inflation,  was necessary to consider 
 the attractors $n_S(N)$ and $r(N)$,
 to rebuild  the
 effective potential and the dissipation coefficient  as a function of the scalar 
 field, respectively
  \cite{Herrera:2018cgi}. Similarly, for the reconstruction of $G-$inflation, was 
  required the spectral index $n_S(N)$ together with the tensor to scalar ratio 
  $r(N)$, in order to reconstruct the potential and the coupling parameter in terms of the inflaton field, 
  see ref.\cite{Herrera:2018mvo}.

Additionally, we mention that in the literature is possible to find 
another methodologies  
to rebuild  the variables as  the scalar potential, scalar spectral index
and the tensor to scalar ratio under the slow roll approximation.
For example, we have the parametrization of  the slow-roll
parameter $\epsilon(N)$, in terms of the  number of $e$-folds $N$
\cite{Huang:2007qz,M,Gao:2017owg}.  Similarly, the reconstruction of the scalar 
potential and spectral index 
from 
 two slow-roll parameters $\epsilon(N)$ and $\eta(N)$,  
was studied 
in ref.\cite{Roest:2013fha}. Also, the reconstruction of the scalar potential, considering as 
ansatz the velocity of the scalar field as a function of the number $N$, in 
 a model of
 $k-$essence inflation  was developed in ref.\cite{Sebastiani:2017cey}.
For other reconstruction methodologies in the scenario of inflation, see  refs.\cite{N1,
N2,yo8}. 

%{\bf ACA VOY}

%Swampland criteria....

%%%%%%%%%%%%%%%%%%%%

On the other hand, in the context of the theoretical foundations
 of  the 
early and present  universe from an effective field theory, there are some criteria or conjectures that have emerged  
 recently in the literature.  These criteria are 
  related  with  the consistency between the 
 effective field theory and superstring theory, in order to describe the universe from
 one
 or various scalar  fields. 
 In this sense, we have
 the Swampland Criteria or Conjectures  (SC)\cite{Sw1,Sw2} and are related to the 
 conditions 
on the range of inflaton field during its dynamic evolution and  also on  the effective potential
(derivatives) associated 
to inflaton field, in order to permit  an 
embedding in the framework of superstring theory\cite{Sw1}. 
This first criterion establishes that the range of inflaton field values $\Delta\phi$ is smaller than 
 the
 Planckian scale during the dynamic of the inflationary epoch. This 
 first conjecture  supposes that 
  the effective field theory  is consistent  with
 the string theory, if  the  range of inflaton field values satisfies  $\Delta\phi<\Delta\, M_p$, 
 where $\Delta$ denotes  a 
 constant of the order $\mathcal{O}(1)$ and $M_p$ denotes the Planck mass\cite{Sw1}. 
 In relation to the condition on the effective potential and its derivatives, we have that the slope 
 of the potential has to be larger to explain that the  fields   coming from the frame 
 of string theory, see refs.\cite{Sw1,Sw2}. Thus, 
  the condition on the slope of the effective potential $V(\phi)$
 (called the second swampland conjecture) can be 
 written as $V_{\phi}/V>c/M_p$, where $V_\phi=\partial V/\partial \phi$ and $c$ 
denotes another constant of the order one as $\Delta$. Additionally, we can 
 have  that the above condition can not be satisfied when the fields are around 
 of the maximum (local) of  the potential, with which we can also consider that 
 $V_{\phi\phi}/V<-c_1/M_p^2$, in which  $c_1$ denotes another constant 
 of the order one\cite{Sw3}. However, we mention that  recently was shown that these constants 
 may be somewhat less than unity, see e.g., ref.\cite{Sw4}. 
 In this sense, under  the theoretical description of inflation in the framework of GR, 
 we can find a direct tension 
 from the second SC and the utilization of 
 the slow-roll approximation, since the slow roll parameter
 $\epsilon\propto (V_\phi /V)^2$   must be smaller than one  during
 inflation i.e., $\epsilon\ll 1$.   
In this way, the imposed conditions by the SC have questioned   whether slow-roll inflation 
described by an effective field  theory.

 In this respect, we mention that the SC do not exclude all inflationary models 
 in the context of the slow roll approximation. In order to describe inflation
 under the slow-roll 
 approximation,
 we have  some  models that can survive to the  requirements 
 of the SC.  In particular, we can mention that for the case of a single scalar 
 field, the model of warm inflation satisfies the criteria imposed by  
 SC\cite{SW6}, see
 also ref.\cite{SW7}.  Also, the SC for a  single field with a 
 chaotic potential in the framework of brane inflation was developed in ref.\cite{SW8}, 
 and this model showed to be  compatible with
  the SC, see also ref.\cite{SW9} for another potentials.  
In the case of quintessential brane inflation and its compatibility with the SC 
introducing  deviations from the Bunch-Davies initial state
was studied in \cite{SW10}.  A curvaton like mechanism is another possibility 
used  in order to conciliate the SC from a single field\cite{SW11} and for multi-field models 
and its compatibly with the 
SC was analyzed in ref.\cite{SW12}.

 Additionally, we comment that  another 
conjecture studied in the literature  is known as 
the Trans Planckian Censorship Conjecture (TCC)\cite{Tp}, 
see also ref.\cite{Tp1}. The TCC  is established
 on the concept that in a suitable  
 quantum theory of gravity  the sub-Planckian quantum 
 fluctuations should persist on a  quantum scale and never become larger 
 than the Hubble horizon and then these fluctuations never become
   freeze during the expansion of the  universe, see also refs.\cite{Tp2,Tp3}.

%%%%%%%%%%%%%%%%%%%%%%%%%%%%%%%%%%%%%%%%%%%%%%%

%{\bf LISTO}

The goal of this investigation is to reconstruct the k-essence inflationary  model, 
considering  the unification between  the attractor or parametrization of
 the scalar spectral index and the slow roll parameter as a  function of the e-foldings together with the 
 SC. 
In this
context, we investigate  how the k-essence  inflationary  model, in which the 
Lagrangian density
 ${\cal{L}}(\phi,X)$ with a new term
 given by
$L(\phi)\,X$ 
 modifies  the
reconstructions of the background variables such as; 
 the scalar potential $V(\phi)$ and
 the coupling parameter $L(\phi)$ and simultaneously satisfy the SC.
 In this sense, we will determine  the structure of
the coupling parameter $L(\phi)$ and of the effective potential $V(\phi)$, in order to
satisfy the swampland criteria and  also the attractor  point associated to the 
scalar spectral index
  $n_S$ from the observations.

In order to satisfy the observational data and the swampland criteria, we consider 
  a general formalism  to rebuild  the effective potential $V$ and the coupling
  parameter $L$,
   from the parametrization of the cosmological
attractor $n_S(N)$ and the slow roll parameter $\epsilon(N)$,
 under   the slow roll approximation.

As an application to the   developed formalism, we will study different  examples 
in order to analyze  the  SC considering   the slow roll 
parameter $\epsilon(N)$ and  also  assuming  the simplest attractor point for the scalar 
spectral index  $n_S-1=-2/N$.
 In this respect,   we will reconstruct the
 effective  potential $V(\phi)$ and the coupling parameter $L(\phi)$ as a 
 function of the inflaton field 
 $\phi$. Additionally, we
 will find 
 different
 constraints on the  parameters in our k-essence model from the unification of 
 the 
  observational 
 data and  the  SC.

%%%%%%%%%%%%%%%%%%%%%%%%%%%%%

The outline of the paper is as follows: The  section II we give
 a brief description  of the model of  k-essence.  The background
 equation and cosmological perturbations are shown. 
 In the section III, we elaborate a general formalism   to rebuild the scalar
 potential and coupling parameter in terms of   the observable  or attractor $n_S(N)$
 and the slow roll parameter. Later in section IV,
we apply the methodology for different   examples in order  to obtain 
the effective potential $V(\phi)$ and the coupling parameter $L(\phi)$,
 as a function of the scalar field $\phi$. 
 In the end, in section V  we give  our conclusions.
 We chose units so that
$c=\hbar=M_p=8\pi=1$.

\section{ The k-essence model }

As  a brief  description of the scenario of k-essence model,  we begin  with
 the 4-dimensional action $S$ for this theory  given by \cite{P1,P2}
\begin{equation}
S = \int \sqrt{-g_{4}}\,d^{4}x\,\left(\frac{1}{2}R
+{\cal{L}}(\phi,X)\right),\label{action}
\end{equation}
where  $g_{4}$ corresponds to  the determinant of the space-time metric
$g_{\mu\nu}$,  $R$ denotes  the Ricci scalar and the quantity ${\cal{L}}(\phi,X)$
represents to the Lagrangian density associated to  the scalar field $\phi$ and $X$. Here the
 quantity $X$ 
corresponds to 
the kinetic energy of the field $\phi$ defined as 
$X=-g^{\mu\nu}\partial_{\mu}\phi\partial_{\nu}\phi/2$.

By   assuming that the energy momentum corresponds to a perfect fluid, 
then it is possible to  identify from the action (\ref{action}) that the energy
density $\rho$ and the pressure $p$ associated to the scalar field $\phi$ and $X$ are
given by \cite{P1,P2}

\begin{equation}
\rho(\phi,X) = 2X\,\frac{\partial {\cal{L}}(\phi,X)}{\partial \,X}-{\cal{L}}(\phi,X),\label{r1}
\end{equation}
and
\begin{eqnarray}
p(\phi,X) = {\cal{L}}(\phi,X),\label{pp1}
\end{eqnarray}
respectively. 
In particular, for the specific   case  in which the Lagrangian ${\cal{L}}(\phi,X)=X-V(\phi)$,
 we recovered the expressions 
 for the energy density and pressure in the framework 
  of the GR. Here the quantity  $V(\phi)$ 
denotes 
 the effective potential associated to scalar field $\phi$.

In this context and  in order to develop the reconstruction for the k-essence model of inflation,
 we will study  the specific  case  in which  the Lagrangian density ${\cal{L}}(\phi,X)$ is given 
 by\cite{P1,P2}
 \begin{equation}
   {\cal{L}}(\phi,X)=X+2\,L(\phi)\,X-V(\phi),\label{anz}
 \end{equation}
where $L(\phi)$ is a coupling  function that depends exclusively on the scalar
field $\phi$. 
We note that in the limit in which this coupling 
parameter $L(\phi)\rightarrow 0$, we recovered the standard GR.

 To analyze this inflationary model,  we consider 
 a spatially  flat  Friedmann Robertson Walker
(FRW) metric, together with  a scalar field homogeneous, such that the field
$\phi=\phi(t)$. In this sense, we have that  the    Friedmann equation can be written as
\begin{equation}
3H^{2} = \,\rho ,\label{HC}
\end{equation}
where $H=\frac{\dot{a}}{a}$ denotes     the Hubble rate and the quantity 
$a$ represents to the scale  factor.  In the following,  the dots denote
differentiation with  respect to the time.

Thus, from relations  (\ref{r1}),  (\ref{pp1}) and (\ref{anz}), we can rewrite 
 the continuity  equation for the  perfect fluid 
 $ \dot{\rho}+3\,H\,(\rho+p)=0,$ as
 \begin{equation}
\ddot{\phi}+3H\dot{\phi}+\frac{V_\phi+L_\phi\,\dot{\phi}^2}{1+2L}=0,\label{cont}
\end{equation}
and also the eq.(\ref{HC}) can be rewrite as  
\begin{equation}
3H^2=\frac{\dot{\phi}^2}{2}+V(\phi)+L\,\dot{\phi}^2.\label{H2}
\end{equation}
Additionally,  combining eqs.(\ref{cont}) and (\ref{H2}) we have 
\begin{equation}
2\dot{H}+3H^2+\frac{1}{2}\dot{\phi}^2-V(\phi)+L\,\dot{\phi}^2=0.\label{DH}
\end{equation}
 In the following, we will assume  that the notation  $V_\phi=\partial V/\partial\phi$,  $L_\phi$  
 corresponds to
  $L_\phi=\partial L/\partial\phi$, $V_{\phi\phi}$ to
 $V_{\phi\phi}=\partial ^2V/\partial \phi^2$,
   etc.

  Introducing  the slow-roll parameters $\epsilon_1$, $\epsilon_2$, $\epsilon_3$  and
$\epsilon_4$ defined as \cite{P1,P2}
\begin{equation}
\epsilon_1=-\frac{\dot{H}}{H^2},\,\,\,\epsilon_2=-\frac{\ddot{\phi}}{H\dot{\phi}},
\,\,\,\epsilon_3=-\frac{L\,X}{H^2},\,\,\,\,\epsilon_4=-\frac{2\,L_\phi\,X}{V_{\phi}},\label{pr}
\end{equation}
and  assuming  that the slow-roll parameters $\epsilon_1$ 
$|\epsilon_2|$, $|\epsilon_3|$, $|\epsilon_4|\ll 1$, during the inflationary regime, 
then   the
   Friedmann equation (\ref{HC}) reduces to \cite{P1,P2}
\begin{equation}
3H^{2}\approx\,\,
V(\phi),\label{HH}
\end{equation}
and the eq.(\ref{cont})
results
\begin{equation}
3H\dot{\phi}\,(1+2L)\approx\,
-V_\phi.\label{eqf}
\end{equation}
Note that combining eqs.(\ref{H2}) and (\ref{DH}) we have that the first slow roll parameter $\epsilon_1$ 
can be rewritten as $\epsilon_1=X(1+2L)/H^2$.
%Here, we have used  the functions given by eq.(\ref{anz}) and the slow roll parameters given by
%Eq.(\ref{pr}).

%Note  that from the slow roll equation
%(\ref{eqf}), we have two limiting situations. The case in
%which the quantity $|2L|\ll 1$, then  it agrees  with slow roll approximation 
%in the framework of the GR for eq.(\ref{eqf}).

% Instead, the inverse case in
%which $ |{\cal{A}}|\gg 1$, the Galileon effect changes the dynamic
%equation of the scalar field $\phi$ and hence the dynamics of
%inflationary model. In this context, we are interested in the latter situation
%in which  the Galileon effect modifies the  dynamics of the G-model and its reconstruction. Thus,
%$3H\dot{\phi}{\cal{A}}\simeq -V_\phi$ and then
%$9H^2\dot{\phi}^2\simeq (V_\phi/g) $ suggesting  that the ratio
%$(V_\phi/g)>0$. Therefore, in the case in which $V_\phi>0$ then
%the quantity $g>0$ and vice versa. In the following we shall take
%$V_\phi>0$ and $g > 0$.

 %Typically, if the scalar field roll down
%potential, then the velocity of the scalar field can be written as
%\begin{equation}
%\dot{\phi}\simeq -\sqrt{\frac{V_\phi}{3\kappa\, g\,V}}.
%\end{equation}
%Here, we have considered Eq.(\ref{HH}). Also, we note that the
%parameter ${\cal{A}}>0$, since we have assumed that $\dot{\phi}<0$.

In general, 
in order to give
 a measure of the inflationary expansion in an inflationary model, we can  define 
 the number of $e$-folds 
$N$ as 
\begin{equation}
N=\int_t^{t_e}H\,dt'=\int_\phi^{\phi_e}H\,\frac{d\phi'}{\dot{\phi}}\simeq
\,\int_{\phi_e}^{\phi}\,V\,\left[\frac{1+2L}{V_{\phi'}}\right]\,d\phi',\label{3N}
\end{equation}
where the quantities  $t$ and $t_e$ correspond to two
different values of cosmological time in which the time $t_e$ indicates   the end of inflationary 
stage and additionally  we have assumed that the number of $e-$folds 
 at the end of inflation is defined as $N(t=t_{e})=0$.

On the other hand, in the context of the cosmological perturbations, the  action 
for the curvature perturbation $\zeta$ for the k-essence model can be 
written as \cite{P1,DeFelice:2013ar,Kobayashi:2010cm}
\begin{equation}
  S^{(2)}=\frac{1}{2}\int\,d\tau 
  d^3x\,z^2[{\cal{G}}(\zeta')^2-{\cal{F}}(\vec{\bigtriangledown}\zeta)^2],\label{ac2}
\end{equation}
where the quantity ${\cal{G}}={\cal{F}}=1+2L$ and the variable $z$ is defined as 
$z=a\dot{\phi}/H$. Here the prime corresponds to derivative with respect to the 
conformal time $\eta=\int dt/a$ and from the Lagrangian density given by  eq.(\ref{anz}),
 we note that the speed of sound associated to 
perturbations $c_S^2=p_X/\rho_X=1$.

From eq.(\ref{ac2}), the scalar  power spectrum of the primordial curvature perturbation is given 
by \cite{P1,DeFelice:2013ar,Kobayashi:2010cm}

\begin{equation}
{\cal{P_S}}=\frac{H^4}{4\pi^2\dot{\phi}^2\,(1+2L)}
\simeq\frac{V^3}{12\pi^2\,V_\phi^2}\,(1+2L).\label{P1}
\end{equation}

Since the scalar spectral index $n_S$ is defined in terms of the power spectrum $\cal{P_S}$ 
as 
$n_S=d\ln{\cal{P_S}}/d\ln k$, we have that  the index
$n_S$ as function  of the standard slow roll parameters $\epsilon$ and $\eta$ can be
written as\cite{P1,DeFelice:2013ar,Kobayashi:2010cm}

\begin{equation}
n_S-1\simeq\,\frac{1}{1+2L}\,\left[2\eta-6\epsilon+\frac{2L_\phi}{1+2L}\sqrt{2\epsilon}\right]
,\label{ns0}
\end{equation}
where the standard parameters $\epsilon$ and $\eta$ are defined by  
\begin{equation}
  \epsilon=\frac{1}{2}\left(\frac{V_{\phi}}{V}\right)^2,\;\,\,
\,\,\,\mbox{and}\,\,\,\eta=\frac{V_{\phi\phi}}{V}.\label{CTM}
\end{equation}

Note that in the limit in which the coupling parameter $L\rightarrow 0$, the spectral index $n_S$ given by
Eq.(\ref{ns0}) reduces to the standard spectral index of the GR, where  $n_S-1\simeq
2\eta-6\epsilon$. Additionally, we have that in the context of the slow roll 
approximation, the relation between the parameters $\epsilon_1$ and $\epsilon$ is 
given by
\begin{equation}
\epsilon_1=-\frac{\dot{H}}{H^2}\simeq\frac{1}{1+2L}\,\epsilon.\label{e11}
\end{equation}

For the case of  the tensorial perturbation, the amplitude
 of the tensor mode in the k-essence model of inflation is  not modified, and its expression 
  is equivalent
  to the standard GR, 
  where 
  the tensor spectrum ${\cal{P_T}}$ is defined as 
${\cal{P_T}}\simeq (H^2/2\pi^2)$. Thus,  we have that the tensor
to scalar ratio $r$
 in the
framework of  k-essence model  can be written as \cite{P1,DeFelice:2013ar,Kobayashi:2010cm}
\begin{equation}
r=\frac{{\cal{P_T}}}{{\cal{P_S}}}=\frac{16X(1+2L)}{H^2}\,=\,16\epsilon_1.\label{rl}
\end{equation}
%One again, note that in the limit ${\cal{A}}\rightarrow 0$ ( or
%equivalently $g\rightarrow 0$), the ratio $r$ coincides with that
%corresponding to GR in which  $r=16\epsilon$.
Also, we can obtain  that the tensor to scalar ratio can be rewritten in terms of 
the standard slow roll parameter $\epsilon$ as
\begin{equation}
r=\frac{{\cal{P_T}}}{{\cal{P_S}}}=\frac{16X(1+2L)}{H^2}\,=\,\frac{16}{1+2\,L}\;\epsilon.\label{rle}
\end{equation}
Here we have considered eqs.(\ref{e11}) and (\ref{rl}), respectively.

In the following we will study the reconstruction of the background variables 
from the unification of the observables parameters  together with  the 
 swampland conjectures or criteria.

%In order to realize the reconstruction we will assume
% an attractor point from the   index $n_S(N)$ and
%the ratio $r(N)$ on the $r-n_S$ plane.

\section{Reconstructing  k-essence model \label{secti2a} }
In this section we will  consider  the methodology  used to
reconstruct  the 
background variables, such as; 
the scalar potential $V(\phi)$ and the coupling  parameter
$L(\phi)$, from the attractor $n_S(N)$ and the slow roll parameter $\epsilon(N)$, together   with  the swampland 
conjectures.

In the context of the reconstruction,
we rewrite    the slow-roll parameters,   spectral index and the tensor to
scalar ratio  as a  function  of the number of $e$-foldings $N$ and its derivatives
\cite{Chiba:2015zpa}.
 From these
expressions    and considering an attractor point  $n_S=n_S(N)$ together with a  parameter 
$\epsilon(N)$,  we should obtain
the  effective potential $V$ and the coupling parameter  $L$ in terms of 
 the number $N$ i.e., $V=V(N)$ and  $L=L(N)$. Now, from 
 Eq.(\ref{3N}) we should
find analytically  the number of  $e$-folds $N$ as a function  of the inflaton field $\phi $ 
and then 
 we should  reconstruct  the scalar potential $V(\phi)$ and the coupling function $L(\phi)$.

In this framework, we start   rewriting
 the derivatives of the potential $V$, coupling function $L$  and  the    slow roll parameters   
in terms of    the number $N$,
as 
$$
V_{\phi}=\frac{dV}{d\phi}=  \,\frac{V\,(1+2L)}{V_{\phi}}
\,V_{\,N}\,,
$$
and then we  get
\begin{equation}
V_{\phi}^2=[V (1+2L)\,\,V_{\,N}]\,.\label{dV}
\end{equation}
In the following, we will assume  that $V_N>0$ and then  the quantity  $1+2L>0$. Also, in the following  
  we will consider 
 the notation  $V_{\,N}=dV/dN$,
 $V_{NN}=d^2V/dN^2$, $L_N$ to $L_{N}=dL/dN$ etc.

For the quantity  $V_{\phi\phi}$, we have 
\begin{equation}
V_{\phi\phi}=\frac{1}{2\,V_N}\,\,\left[(1+2L)\,[V_N^2+V\,V_{NN}]+2L_N\,V\,V_N\right],\label{ddV}
\end{equation}
and for the coupling parameter $L_\phi$, we get
\begin{equation}
L_\phi=\sqrt{\frac{V(1+2L)}{V_N}}\,L_N.
\end{equation}

From these relations, we find that  the standard slow roll parameters $\epsilon$ and
$\eta$  given by eq.(\ref{CTM}) can be  rewritten as 
\begin{equation}
\epsilon=\frac{1}{2}\,(1+2L)\,\frac{V_N}{V},\label{p1}
\end{equation}
and
\begin{equation}
\eta=\frac{(1+2L)}{2}\,\left[\frac{V_N}{V}+\frac{V_{NN}}{V_N}\right]+L_N
,\label{p2}
\end{equation}
respectively.

Also, from eq.(\ref{3N}) we obtain that 
 the relationship  between the $e$-folding $N$ and the inflaton
 $\phi$  becomes
 \begin{equation}
   \int
   \left[\frac{V_{\,N}}{(1+2L)\,V}\right]^{1/2}\,dN=\int\,d\phi.\label{NF}
 \end{equation}

In relation to the observables, we find that the power spectrum of the primordial curvature 
perturbation $\cal{P_S}$ as a function of the number of e-folds  becomes
 \begin{equation}
   {\cal{P_S}}=\frac{1}{12\pi^2}\,\frac{V^2}{V_N},\label{PN1}
 \end{equation}
and as we can note this quantity does not depend on the coupling parameter $L$, under this 
formalism.

Also, from   eqs.(\ref{ns0}),
(\ref{p1}) and (\ref{p2}), we obtain that  the scalar spectral index $n_S$ can be rewritten in
terms of the $e$-folds  $N$ as 
\begin{equation}
n_S-1=-2\frac{V_{\,N}}{V}+\frac{V_{\,NN}}{V_{\,N}}=\left[ \ln\left(\frac{V_N}{V^2}\right)\right]_N,
\label{NSN1}
\end{equation}
and 
by considering  Eq.(\ref{rl})    the tensor to scalar ratio can be rewritten as 
\begin{equation}
  r=\frac{{\cal{P_T}}}{{\cal{P_S}}}=8\,\frac{V_{\,N}}{V}.\label{gr8}
\end{equation}

 Here, we note that  the expressions for the power spectrum ${\cal{P_S}}(N)$,  the scalar spectral index $n_S(N)$ and 
 the tensor to scalar ratio  $r(N)$ 
 are the same 
  to the one obtained in the standard  GR \cite{Chiba:2015zpa} and these observables depend 
  exclusively on the effective  potential and its derivative with respect to $N$. This suggests 
  that the coupling parameter $L(\phi)$ can not be rebuild from the one 
  attractor, such as; the scalar spectrum or the scalar  spectral index $n_S(N)$ or from the tensor to 
  scalar ratio $r(N)$, as can be seen from eq.(\ref{PN1}) or (\ref{NSN1}) or (\ref{gr8}).
However, we observe from eq.(\ref{NF}) that the relation between the number of $e$-folds and the 
inflaton field depends on the coupling parameter $L$ as well as $V$, and then in 
order to reconstruct the background in our model, we should  know  the function 
$L(N)$.

Thus, as in the case of GR the effective potential $V$ in terms of the number $N$ 
can be obtained from the attractor $n_S(N)$ as 
\begin{equation}
V(N)=-1/\int\left(\exp\left[\int(n_S-1)dN\right]\right)dN,\label{VNN}
\end{equation}
or also giving the attractor $r(N)$ with which we have that the effective 
potential becomes
\begin{equation}
V(N)=\,\exp\left[\frac{1}{8}\int\,r\,dN\right].\label{Vw}
\end{equation}
 
Here, we emphasize  that the reconstruction of the effective potential  in terms of the scalar field $\phi$
i.e., $V(\phi)$ can not 
be determined without the help of 
the coupling parameter $L(N)$, see eq.(\ref{NF}). In this sense, we make the difference with 
respect to the standard GR, although the expressions for the observables 
parameters 
 $n_S(N)$ and $r(N)$ are the same to GR, as can be seen of eqs.(\ref{NSN1}) and (\ref{gr8}).

Thus, a possible solution to rebuild the background variables can be through an ansatz 
on the slow roll parameter $\epsilon$ in terms of the number $N$ i.e., $\epsilon(N)$. 
This ansatz on the slow roll parameter 
$\epsilon$ would have to 
 take into account 
the swampland conjecture in which $V_\phi/V\sim \mathcal{O}(1)$, since this 
parameter $\epsilon\propto (V_\phi/V)^2$. In this respect, the coupling parameter $L(N)$ can be 
determined considering eq.(\ref{p1}) in which
\begin{equation}
L(N)=\left[\frac{\epsilon\,V}{V_N}-\frac{1}{2}\right],\label{LN}
\end{equation}
where  the slow roll parameter  $\epsilon=\epsilon(N)$.

Note that the relation between the number $N$ and the inflaton $\phi$ given by eq.(\ref{NF}) can be 
rewritten as
\begin{equation}
\int\,\frac{V_N}{V}\,\left[\frac{1}{2\epsilon}\right]^{1/2}\,\,dN=\int\,d\phi.\label{NPG}
\end{equation}
This suggests that first swampland criterion on the field range traversed by 
scalar field during the slow roll epoch in an effective field theory $\Delta\phi<\Delta\sim  \mathcal{O}(1)$ can be written as
\begin{equation} 
  \phi-\phi_0=\Delta\phi=\int_{N_0}^N\,\frac{V_{N'}}{V}\,
  \left[\frac{1}{2\epsilon}\right]^{1/2}\,\,dN'=\int_{N_0}^N\,
  \sqrt{\frac{r}{8(1+2L)}}\,dN'
  <\Delta\sim  \mathcal{O}(1),\label{DF1}
\end{equation}
where $N(\phi=\phi_0)=N_0$ corresponds to the number of $e-$folds 
during the slow roll epoch and its value is such that 
$0<N_0<N$. Recalled that during the slow roll scenario we can consider that the number of $e-$folds 
is large  ($\mathcal{O}(10)\sim\mathcal{O}(10^2)$), see 
ref.\cite{Chiba:2015zpa,Herrera:2018cgi}. Also,
we note that in the special case in which the ratio $r$ and $L$ are constants, then we
 have the  relation 
$\Delta\phi\propto \Delta N<\Delta$, in which $\Delta N=N-N_0>0$.

In the context of  the second swampland conjecture we consider that  
 this  criterion  is related to the 
slope of the effective  potential, then we can associate the parameter $\epsilon$ 
to this conjecture, such as, 
the parameter $\sqrt{2\epsilon}=(V_\phi/V)> c$, where the constant $c$ from the swampland conjecture is  
of 
order one, i.e., 
 $c\sim{ \cal{O}}(1)$ (in Planck unit) as we mentioned before.

In this sense,  from eqs.(\ref{rle}) and (\ref{gr8}) (or eq.(\ref{p1})), we have 
\begin{equation}
  (1+2L)=2\,\epsilon\,\left(\frac{V}{V_N}\right),
\end{equation}
and considering the second swampland conjecture in which $\epsilon>c^2/2 $, we 
obtain a lower bound for the coupling parameter $L$ given by
\begin{equation}
  L>\,c^2\,\left(\frac{V}{2\,V_N}\right)-\frac{1}{2}.
\end{equation}
In this form, we can obtain some constraints on the parameters-space from the swampland criteria 
in order to 
rebuild  the k-essence model.

In the following, under the slow roll approximation we will study some specific examples 
in order to reconstruct
the scalar potential $V(\phi)$ and coupling parameter $L(\phi)$, from the
attractor $n_S(N)$ and  $\epsilon(N)$   together with the swampland criteria.

\section{Unifying swampland criteria  and  attractor point }

In this section we will apply 
 the formalism of above in order to rebuild the background variables in our k-essence model. In this context,  
  we shall use  the simplest
 attractor  for the scalar spectral index  $n_S(N)$ together with some examples 
 for the slow roll parameter $\epsilon(N)$,
 in order to find analytically the effective potential $V(\phi)$ and the coupling parameter $L(\phi)$ 
 in terms of the inflaton field $\phi$.
 Thus, 
 following
refs.\cite{T,Chiba:2015zpa} we consider  that for large $N$ (slow roll regime), the simplest  attractor point 
for the scalar spectral
index in terms of the number of $e$-foldings   can be written as 
\begin{equation}
n_S-1=-\frac{2}{N},\label{e1}
\end{equation}
in which for $N=60$ the scalar spectral index is well corroborated from the Planck data. 
As we mentioned before, here large $N$ corresponds to values of the number of 
$e-$foldings 
$\sim{\cal{O}}(10)\sim {\cal{O}}(10^2)$, during the slow roll stage
 \cite{T,Chiba:2015zpa},
 and the point $N=0$ 
is not allowed.  In this sense, this parametrization on the observable $n_S(N)$ does  
not  pretend  to describe the end of the inflation where the number $N=0$, but it characterizes 
the slow roll regime in which the number of $e-$folds $N$ is large.
In the following, we will consider the value of the number of  $e$-folds $N=60$,
in order to evaluate the observational parameters and we will only analyze the 
attractor point $n_S(N)$ given by relation (\ref{e1}), for two different slow roll parameters $\epsilon(N)$.

In this way, replacing  eq.(\ref{e1}) into eq.(\ref{NSN1}) and integrating we have\cite{Chiba:2015zpa} 
$$
\frac{V_N}{V^2}=\frac{\alpha}{N^2},
$$
where the parameter $\alpha$ corresponds to an integration constant and as we 
have assumed that $V_N>0$, then the parameter $\alpha>0$. However, from eq.(\ref{PN1}) we can note 
that this integration constant $\alpha$ can be fixed considering the scalar power 
spectrum $\cal P_S$, when the wavelength of the perturbation crosses the Hubble radius 
( at $N=60$)
  results
 \begin{equation}
   \alpha=\frac{N^2}{12\pi^2\cal{P_S}}.
 \end{equation}
In this sense, for the specific case in which the number $N=60$ and the scalar 
power 
spectrum ${\cal{P_S}}\simeq 2.2\times 10^{-9}$, we have that the value of $\alpha\simeq 
10^{10}$.

Now, from eq.(\ref{VNN}) 
the effective potential in terms of the number of $e$-folds $N$ can be written 
as
\begin{equation}
V(N)=\frac{N}{\alpha+\beta\,N},\label{V1}
\end{equation}
where the quantity $\beta$ corresponds to a second integration constant and rigorously this 
constant can be chosen to be  positive, negative or zero. However, we can obtain 
an estimate of this constant considering the tensor to scalar ratio (\ref{gr8}) in which
 \begin{equation}
   \beta=\frac{\alpha}{N}\left[\frac{8}{Nr}-1\right],\,\,\,\mbox{for}\;\;\beta>0,\,\,\,\mbox{and}\,\,\;\,\,\;
 \beta=\frac{\alpha}{N}\left[1-\frac{8}{Nr}\right],\,\,\,\mbox{for}\;\;\beta<0.\label{BET}
 \end{equation}
Thus,  for the case in which $\beta>0$ and using the values $N=60$, $\alpha=10^{10}$ and considering 
that the tensor to 
scalar ratio $r<0.07$ from observational data, we obtain a lower limit for   
the integration constant $\beta$   given by   
$\beta>1.5\times10^{8}$. For the case in which the integration constant 
 $\beta$ is negative, we find a upper bound given by $\beta<-1.5\times 10^8$.

In order to rebuild the effective potential $V(\phi)$ and the coupling parameter $L(\phi)$, 
we should  know the 
parameter $L(N)$ to perform the integral given by eq.(\ref{NF}) i.e., we  should find the relation between 
the number of $e$-folds and the inflaton field ($N=N(\phi)$) for the reconstruction 
of the functions $V(\phi)$ and $L(\phi)$.

In fact, to find the coupling parameter $L(N)$ and then $N=N(\phi)$, we can consider 
an ansatz on 
the slow roll parameter $\epsilon=\epsilon(N)$ together with the attractor point $n_S(N)$ 
from the relation given by eq.(\ref{LN}). The motivation to consider  
 this ansatz on the 
slow roll parameter $\epsilon(N)$ is associated to satisfy  the second  
swampland conjecture, since the parameter $\epsilon$ is proportional to  $ (V_\phi/V)^2$.  

As a first 
ansatz on the variable $\epsilon(N)$, we consider the simplest situation in which the slow 
roll parameter $\epsilon$ is equal to a constant 
\begin{equation}
\epsilon(N)=\epsilon_0=cte.
\end{equation}
In order to satisfy the second swampland conjecture, we can assume that this constant  $\epsilon_0 $ 
satisfies the lower bound  $\epsilon_0> c^2/2 \sim1/2$. 

Immediately, we can see that from eq.(\ref{CTM}) and  this ansatz 
for the slow roll parameter,  the effective potential as a function of the 
inflaton field corresponds to an exponential potential with which 
\begin{equation}
V(\phi)=\exp[A_1\phi+C]=\exp[\,\tilde{\phi}\,],\label{POT1}
\end{equation}
where  the field $\tilde{\phi}$ is defined as $\tilde{\phi}=A_1\phi+C$, in which 
the quantity 
$A_1=\sqrt{2\epsilon_0}$ and $C$ corresponds to an integration 
constant. Thus, we note that the attractor given by eq.(\ref{e1}) is not necessary  
for the reconstruction of the $V(\phi)$. However, in order to rebuild the 
coupling parameter $L(\phi)$, we need an ansatz on the attractor point $n_S(N)$. 

In this context, from eq.(\ref{LN})
 we  obtain the 
the coupling parameter $L(N)$ in terms of the number of e-folds $N$ becomes
 \begin{equation}
L(N)=\,\frac{\epsilon_0\,N(\alpha+\beta\, N)}{\alpha}-\frac{1}{2}.\label{LN1}
\end{equation}
Here we have used the relation for the potential $V(N)$ given by  eq.(\ref{V1}) obtained 
from attractor $n_S(N)$ given by  eq.(\ref{e1}). 

%Thus, from the second swampland conjecture we can consider 
%that the lower limit for the coupling quantity becomes
 %\begin{equation}
  % L(N)> \,\frac{c^2\,N(\alpha+\beta\, N)}{2\,\alpha}-\frac{1}{2}.
 %\end{equation}

Now, considering the relation between the number $N$
and the scalar field given by eq.(\ref{NPG}) and assuming the special case in which $\beta>0$, 
we have that the number $N(\phi)$ is given by  
 \begin{equation}
N(\phi)=\frac{\alpha\,\exp[\,\tilde{\phi}]}{\alpha-\beta\,\exp[\,\tilde{\phi}]}.\label{NP1}
 \end{equation}
Here we can note that replacing eq.(\ref{NP1}) into eq.(\ref{V1}) we recover the effective 
potential $V(\phi)=e^{\tilde{\phi}}$ given by eq.(\ref{POT1}). Also, we 
observe that number of $e$-folds has a pole at $\tilde{\phi}=\ln(\alpha/\beta)$ 
when $\beta>0$,
and then we can only consider that the range for the scalar field $\tilde{\phi}$ 
is given by
$\tilde{\phi}<\ln(\alpha/\beta)$, since the number $N>0$, see eq.(\ref{NP1}).

Additionally, we 
can mention that for the case in which the integration constant $\beta<0$ the 
number $N=N(\phi)$ results
$N(\phi)=\frac{\alpha\,\exp[\,\tilde{\phi}]}{\alpha+(-\beta)\,\exp[\,\tilde{\phi}]}$ and 
this number $N$ does not contain a pole. 

Also, 
for the specific  case in which the constant $\beta=0$, we have that the number 
of $e-$foldings evolves exponentially as 
$N(\phi)=\,e^{\,\tilde{\phi}}$ and we note that the number of $e-$foldings does not depend on the
integration constant $\alpha$. Here, we mention that in this  specific case in which the integration constant 
$\beta=0$, we obtain that the effective potential is reduced to  $V(N)=N/\alpha$ and then  
combining  eqs.(\ref{gr8}) and (\ref{e1}), the consistency relation $r=r(n_S)$ is given by 
$r=4(1-n_S)$. In this sense, for the value $n_S=0.967$ the tensor to scalar ratio $r\simeq 0.13$ and this value 
is disapproved from the Planck data. 
 Furthermore for $\beta=0$, we note that this consistency 
relation $r=4(1-n_S)$ becomes independent on the chosen parameters 
$\epsilon(N)$ and then this rebuilt model does not work.

Thus for $\beta>0$, we find that the reconstruction for the coupling function $L(\phi)$
by combining eqs.(\ref{LN1}) and (\ref{NP1}) becomes
 \begin{equation}
L(\phi)=\frac{\epsilon_0\,\alpha^2\,e^{\,\tilde{\phi}}}{(\alpha-\beta\,e^{\,\tilde{\phi}})^2}-\frac{1}{2}
=\frac{\epsilon_0\,\alpha^2\,V(\phi)}{[\alpha-\beta\,V(\phi)]^2}-\frac{1}{2},\label{R1}
 \end{equation}
with $\alpha>\beta V(\phi)$, since the number of $e-$foldings $N$ is defined as positive.
%In the particular case in which the parameter $\alpha\gg \beta V$, we observe 
%that the coupling function $L(\phi)$  evolves exponentially with the scalar 
%field, since in this limit $L(\phi)\propto V(\phi)\propto e^{\phi}$.

For the case in which the integration constant  $\beta<0$, we have 
\begin{equation}
L(\phi)=\frac{\epsilon_0\,\alpha^2\,e^{\,\tilde{\phi}}}{(\alpha+(-\beta)\,e^{\,\tilde{\phi}})^2}-\frac{1}{2}
=\frac{\epsilon_0\,\alpha^2\,V(\phi)}{[\alpha+(-\beta)\,V(\phi)]^2}-\frac{1}{2}.\label{R2}
 \end{equation}
%Here we also note that  the limit $\alpha\gg \beta V$ we have that the coupling 
%parameter $L(\phi)\propto V(\phi)\propto e^{\phi}$ as before. However, in the 
%case in which the constant $\beta$ is negative  the number $N$ does not contain a pole, with which
%we can consider the opposite limit  $\alpha\ll \beta V$, and here the coupling function 
%evolves as $L(\phi)\propto V(\phi)^{-1}\propto e^{-\phi}$.

For the situation in which the quantity $\beta=0$, we find that the reconstruction for the coupling parameter 
$L(\phi)$ is given by
\begin{equation}
  L(\phi)=\epsilon_0\,\,e^{\,\tilde{\phi}}-\frac{1}{2}.
\end{equation}

In relation to the unification between the observational parameters and the SC, 
we  consider the first swampland conjecture in which $\Delta\phi< 
\Delta\,\sim\mathcal{O}(1)$, for our first example.  In this sense, we can  rewrite eq.(\ref{NP1}) for the case $\beta>0$ in which the integration 
constant $C$ is defined as  $C=\ln[N_0/(\alpha+\beta N_0)]-\sqrt{2\epsilon_0}\phi_0$,
 wherewith the range of the inflaton field during the slow roll regime results
\begin{equation}
  \Delta\phi=\phi-\phi_0=\frac{1}{\sqrt{2\epsilon_0}}\,\ln\left[\frac{N\,(\alpha+\beta N_0)}
  {N_0(\alpha+\beta N)}\right] <\Delta\,\sim 1,
\end{equation}
 as before $N(\phi=\phi_0)=N_0$   
 and its value is $0<N_0<N$,  see eq.(\ref{DF1}).  Thus, from this conjecture and 
 for the case $\beta>0$, we can find  
 a lower bound for the value 
 $\epsilon_0$ given by 
\begin{equation}
  \epsilon_0>\frac{1}{2}\,\,\ln^2\left[\frac{N\,(\alpha+\beta [N-\Delta N])}
  {[N-\Delta N](\alpha+\beta N)}\right],\label{qqq}
\end{equation}
where recalled that the quantity $\Delta N$ is defined as $\Delta N=N-N_0>0$.

 In particular 
for the case in which $\Delta N=50$ (or $N_0=10$) and $\alpha=10^{10}$, 
we find that the integration constant 
$\beta$ becomes $\beta<3.2 \times 10^8$ according eq.(\ref{qqq}) together with
the constraint   $\epsilon_0>c^2/2\sim0.5$ 
(second swampland conjecture). 
However, from the tensor to scalar ratio in which $r<0.07$ at $N=60$ we have found that 
$\beta>1.5\times 10^{8}$, see relation  given by eq.(\ref{BET}). In this form, 
we obtain that
 the range for the parameter $\beta$ unifying 
the observational data together with the swampland conjectures becomes
\begin{equation}
  1.5\times 10^8<\beta<3.2 \times 10^8.\label{co1}
\end{equation}
In this sense, we observe that the range for the parameter $\beta$ is very arrow if we want to satisfy
the observational and theoretical conditions. Similarly, for the situation in which the  integration constant 
$\beta$  is negative, we find that the range for this parameter  is given by 
\begin{equation}
  -1.5\times 10^8>\beta>-3.2 \times 
10^8.\label{co2}
\end{equation}

% \begin{equation}
%   N_0>\alpha\,\left[\frac{(\alpha+\beta 
 %  N)}{N}\,e^{\sqrt{2\epsilon_0}}-\beta\right]^{-1}>0.
% \end{equation}
%In particular for $N=60$, $\alpha=10^{10}$, $\beta=2\times 10^8$ and $\epsilon_0=0.8$ 
%we find that the lower limit for the number $N_0$ is given by $N_0>9$.

As a second 
ansatz on the variable $\epsilon(N)$, we can assume 
that for large N, the slow 
roll parameter $\epsilon(N)$ can be considered as \cite{Huang:2007qz,Gao:2017owg,Gron:2018rtj}
\begin{equation}
\epsilon(N)=\frac{\gamma}{N},\label{AS2}
\end{equation}
where $\gamma$ corresponds to a constant.  We note that from ref.\cite{Gao:2017owg},  we can recognize that for 
large $N$, the
constant $\gamma$ is equal to  $\gamma=1/4$ for this parametrization on $\epsilon(N)$. 
However,  applying  the second swampland conjecture, we have  that 
the slow roll parameter $\epsilon>c^2/2$, then the constant $\gamma$ can  
satisfy the lower bound given by $\gamma>N\,c^2/2$. Thus, for the case 
in which the number of $e$-folds  $N=60$ and $c\sim 1$,  we get that the lower limit for $\gamma$
results $\gamma>30$.

From  this ansatz on $\epsilon(N)$, we find that the coupling parameter $L$ as a 
function of the number of e-folds 
results
 \begin{equation}
   L(N)=\frac{\gamma(\alpha+\beta\,N)}{\alpha}-\frac{1}{2}.\label{38}
 \end{equation}

Now, from  eq.(\ref{NPG}) and assuming the case in which the constant $\beta$ is  positive,  we 
get that the number 
of $e-$foldings in terms of the scalar field becomes 
\begin{equation}
N(\phi)=\frac{\alpha}{\beta}\,\tan^2\left[\frac{1}{2}\;\sqrt{\alpha\;\beta}\,(A_2\,\phi+C)\right],\label{NP2}
\end{equation}
where $A_2$ is defined as $A_2=\sqrt{2\;\gamma}/\alpha$ and $C$ corresponds to 
an  integration constant. Here we note that in order to 
evade the singularity on the number $N=N(\phi)$
given by  eq.(\ref{NP2}),  
 we have  that  $0<\sqrt{\alpha\beta}\,(A_2\phi+C)\lesssim \pi$. 
For the situation in which the integration constant 
$\beta<0$, we obtain that the relation between $N$ and $\phi$ can be written as 
\begin{equation}
N(\phi)=\frac{\alpha}{(-\beta)}\,\tanh^2\left[\frac{1}{2}\;\sqrt{\alpha\;(-\beta)}\,(A_2\,\phi+C)\right].
\end{equation}
In particular for the special case in which $\beta=0$, we find that the relation $N=N(\phi)$ 
is given by 
\begin{equation}
N(\phi)=\frac{1}{4}\,\left[\sqrt{2\,\gamma}\,\phi\,+C\right]^2.
\end{equation}
Again, we note that this relation $N=N(\phi)$ for the case $\beta=0$ does not depend of 
the integration constant $\alpha$. 

Thus, for the case $\beta>0$, we obtain that the reconstruction for the effective potential $V(\phi)$  
combining eqs.(\ref{V1}) and (\ref{NP2}) results
\begin{equation}
V(\phi)= 
\frac{1}{\beta}\,\sin^2\left[\frac{1}{2}\;\sqrt{\alpha\;\beta}\,(A_2\,\phi+C)\right]=\frac{1}{2\,\beta}
\left[1-\cos\left(\sqrt{\alpha\;\beta}\,[A_2\,\phi+C]\right)\right],\label{V22}
\end{equation}
and this effective potential corresponds to  natural inflation\cite{Freese:1990rb,Adams:1992bn}, 
see also \cite{Ni1}. In this framework, the scalar field  is associated to a pseudo Nambu-Goldstone 
boson (pNGb) with a  pNGb potential given by (\ref{V22}). Here the constants $\beta^{-1/4}$ 
and $1/\sqrt{\alpha\beta}\,A_2$ can be associated to the mass scales from the 
particle physics models, for more details see \cite{Freese:1990rb,Adams:1992bn}.

For the case in which the integration constant $\beta$ is negative, we have that the reconstruction of the
effective potential as a function of the scalar field, becomes 
\begin{equation}
V(\phi)= 
\frac{1}{(-\beta)}\,\sinh^2\left[\frac{1}{2}\;\sqrt{\alpha\;(-\beta)}\,(A_2\,\phi+C)\right], \label{V23}
\end{equation}
and this potential corresponds to  Hyperbolic inflation 
\cite{Basilakos:2015sza}. Also,  we mention that this potential has an 
interesting application to describe the dark energy  and
 then the later time acceleration of the present universe\cite{Hp1}.
  Additionally,  we have that this hyperbolic potential has a behavior of an 
exponential or power law potential, depending on the specific limits taken for the scalar field. 
In this way, in the case in which $\lvert\sqrt{\alpha (-\beta)}(A_2\phi+C)/2\lvert\gg 1 $, we 
have an exponential potential $V(\phi)\propto e^{-\sqrt{\alpha (-\beta)}\,A_2\,\phi}$ 
and in the opposite limit  $\lvert\sqrt{\alpha (-\beta)}(A_2\phi+C)/2\lvert\ll 1 $, 
we get a quadratic potential $V(\phi)\propto \phi^2$.

In the special case in which the integration constant $\beta=0$, we obtain that 
the reconstruction for 
the effective potential is given by
\begin{equation}
V(\phi)=\frac{1}{4\alpha}\,\,\left[\sqrt{2\,\gamma}\,\phi\,+C\right]^2,\label{V24}
\end{equation}
and it corresponds to a quadratic  potential (chaotic potential) i.e.,  
$V(\phi)\propto\phi^2$ and it coincides with  the potential given by eq.(\ref{V23}) 
in 
the limit 
$\lvert\sqrt{\alpha (-\beta)}(A_2\phi+C)/2\lvert\ll 1 $.

For the reconstruction of the coupling parameter $L(\phi)$, we can combine eqs.(\ref{38}) 
and (\ref{NP2}) for the case in which $\beta>0$ obtaining 
\begin{equation}
L(\phi)=\gamma\left(1+\tan^2\left[\frac{1}{2}\;\sqrt{\alpha\;\beta}\,(A_2\,\phi+C)\right]\right)
-\frac{1}{2}=\frac{\gamma}{(1-\beta V(\phi))}-\frac{1}{2},\label{LL1}
\end{equation}
and in order to have $1+2L>0$, then it is necessary that during the inflationary epoch $\beta V<1$.

 In the situation in which the integration constant  $\beta$ is  negative we have 
\begin{equation}
L(\phi)=\gamma\left(1+\tanh^2\left[\frac{1}{2}\;\sqrt{\alpha\;(-\beta)}\,(A_2\,\phi+C)\right]\right)
-\frac{1}{2}=\frac{\gamma}{(1+(-\beta) V(\phi))}-\frac{1}{2}.\label{LL2}
\end{equation}
Note that in this case we have the possibility to consider the regime in which 
$1<(-\beta)V(\phi)$, for the function $L(\phi)$ that increases as $L(\phi)\propto 1/V(\phi)$, when 
the potential
decreases during the slow roll scenario. 

Now, for  the situation in which the integration constant $\beta$ is equal to zero, we 
obtain 
that the reconstruction of the  coupling parameter 
$L(\phi)=\gamma-1/2=$ constant and positive, since $\gamma>30$ in order to satisfy the SC.
 Again, we mention that in this  specific case in which the integration constant 
$\beta=0$, 
%we obtain that the effective potential $V(N)=N/\alpha$ and then  
%combining  eqs.(\ref{gr8}) and (\ref{e1}) the consistency relation $r=r(n_S)$ is given by 
%$r=4(1-n_S)$. In this sense, for the value $n_S=0.967$ 
the tensor to scalar ratio $r$ is disapproved from the data 
and then this reconstructed model does not work, since the consistency relation $r=r(n_S)$ 
becomes 
$r=4(1-n_S)$.

%in the  framework of    standard GR (in our case $L=0$), the tensor to scalar ratio 
%$r$ is  defined as $r=16 \epsilon$ and then for 
 %a quadratic potential $V(\phi)\propto\phi^2$, we know that the  tensor to scalar ratio 
%$r$   is disapproved from the Planck data\cite{Planck2018}. 
%However, in 
%the case of k-essence this ratio is modified by the term $(1+2L)^{-1}$ coming from the scalar power 
% spectrum, with which we have  
%$r=(16/[1+2L])\,\epsilon$, see eq.(\ref{rle}). 
%In our k-essence model in
  %the case in which the integration constant 
 %$\beta=0$  the situation 
%becomes similar to standard GR where the ratio $r$ is disapproved form the data.

% in the special situation in 
%which the integration constant $\beta=0$ that gives rise to the reconstruction of 
%a quadratic potential together with a coupling parameter constant the 

As before, in order to consider the first conjecture in the reconstruction of k-essence inflation,
 we rewrite the eq.(\ref{NP2}) 
 %for the case in which 
%t%he constant $\beta>0$ to find a limit on the parameter $\gamma$. In this 
%context, 
and then    the range $\Delta\phi$ in terms of the variation  $\Delta N=N-N_0$, can be written as
\begin{equation}
  \Delta\phi=\frac{2}{\sqrt{\alpha\beta}\,A_2}\,\arctan\left[\sqrt{\frac{\beta\,\Delta 
  N}{\alpha}}\,\right]<\Delta \sim 1.\label{dp2}
\end{equation}
 In this form, from eq.(\ref{dp2}) we 
find that a lower bound for the parameter $\gamma$ given by 
\begin{equation}
\gamma>\left(\frac{2\,\alpha}{\beta}\right)\,\arctan^2\left[\sqrt{\frac{\beta\,\Delta 
  N}{\alpha}}\,\right].\label{ss}
\end{equation}
From the lower bound for $\gamma$ given by eq.(\ref{ss}) and under the condition 
$\gamma>N\,c^2/2\simeq 30$ imposed by the second SC, we find numerically that the upper bound for the
 integration constant $\beta$ is given by $\beta<8.3 \times 10^8$. Here we have 
 used the values $\alpha=10^{10}$ and $\Delta N=50$. 
 
 In this way, we obtain that 
 the range for the integration constant $\beta$ under the unification  of the upper bound for the 
 ratio $r<0.07$ and 
 the swampland criteria is
 \begin{equation}
 1.5\times 10^{8}  <\beta<8.3\times 10^{8}.
 \end{equation}
Thus,   to satisfy the 
observational data together with the swampland criteria, 
we note that this range for the parameter $\beta$ is not that  narrow as the previous case in which 
$\epsilon(N)=\epsilon_0=$ constant.   Also, we note that this result suggests that the mass 
scale $\beta^{-1/4}\sim { \cal{O}}(10^{-2})$ (in units of Planck mass) is 
similar to obtained in the framework of GR, where $\beta^{-1/4}\sim$ GUT scale $(\sim{ \cal{O}}(10^{-4}))$
\cite{Adams:1992bn}.

Analogously, for the situation in which the parameter $\beta$ is negative, we find 
that the range for $\beta$  is given by $-10^8>\beta>-10^{9}$, assuming the values of 
$\alpha=10^{10}$ and the variation $\Delta N=N-N_0=50$.

%number of e-foldings $N_0$ is given by 
%\begin{equation} 
 % N_0>N-\frac{\alpha}{\beta}\,\tan^2\left[\sqrt{\frac{\gamma\;\beta}{2\,\alpha}}\right]>0,
%\end{equation}constant
%with the rate $\gamma\beta/\alpha\neq\pi^2/2,9\pi^2/2,.$ In particular for the 
%values of 
%$N=60$, $\alpha=10^{10}$, $\beta=2\times 10^{8}$ and $\gamma=50$, we find that 
%the lower limit for the number $N_0$ is given by $N_0>23$.

\section{Conclusions \label{conclu}}

 In this article we have investigated the reconstruction of the background 
 variables 
 in  the k-essence inflationary model from the Lagrangian density given by eq.(\ref{anz}). 
 To rebuild this scenario, we have 
  unified the swampland criteria together 
 with the attractor point associated to 
  the scalar
 spectral index
  $n_S(N)$ and the slow roll parameter $\epsilon(N)$, in which $N$ denotes  the
number of $e$-folds. From a coupling of the form $L(\phi)X$ in the k-essence 
model, we have found a general treatment of reconstruction in the framework of 
slow-roll approximation. In this sense, we have obtained integrable results for the scalar potential
$V(N)$ and the coupling parameter $L(N)$, in terms of the scalar spectral index $n_S(N)$ 
and the slow-roll parameter $\epsilon(N)$. Interestingly, we have found that the 
effective potential as a function of the number $N$ i.e., $V(N)$ coincides with 
the expression obtained in the framework of GR, (see eq.(\ref{VNN})) and it does 
not depend of the coupling parameter $L$. However, in order to rebuild the 
effective potential in terms of the scalar field $V(\phi)$, we need to consider 
an expression for the coupling parameter $L$ as a function of the number of 
$e-$foldings $N$ i.e., $L(N)$. In this respect, we make the difference with 
respect to the reconstruction of the effective potential  $V(\phi)$ in the framework of the  GR.
Additionally, in this general analysis we have obtained   from the  
slow roll parameter $\epsilon(N)$  and the attractor point $n_S(N)$ an expression for 
the coupling parameter $L(N)$, see eq.(\ref{LN}). 

In order to apply this  reconstruction methodology during the slow roll regime, we have considered  the simplest 
example for the scalar spectral index $n_S(N)$ given by $n_S=1-2/N$, together 
with two specific ansatze for the slow roll parameter $\epsilon(N)$ under the assumption  of  
 large 
$N$. In this sense, we have utilized  the special cases in which the slow roll 
parameter $\epsilon(N)$ is a constant and when $\epsilon(N)\propto N^{-1} $, in 
order to rebuild the effective potential $V(\phi)$ and the coupling parameter 
$L(\phi)$.

In the case in which the slow roll parameter is a constant, we have found that 
the effective potential as a function of the scalar field  corresponds to an exponential potential, 
as can be seen of eq.(\ref{POT1}).
In this situation,  we have noted that the attractor $n_S(N)$ was not necessary  to rebuild the 
exponential potential $V(\phi)$, since from definition of the slow roll parameter is possible to obtain the
effective potential. In relation to the coupling parameter $L(\phi)$ when $\epsilon(N)=$ constant, 
we have found 
that the reconstruction of this parameter 
is given by   eq.(\ref{R1}) for the case in which the integration constant $\beta$ 
is positive and  for the case in which 
$\beta<0$, the coupling parameter  $L(\phi)$ is given by eq.(\ref{R2}). 
%Here, we  have noted that when $\beta V(\phi)\ll \alpha$,
%the $\beta V(\phi)\gg 1$ the coupling 
%parameter $L(\phi)$ decays as  $L(\phi)\propto 1/V(\phi)$ and in the opposite situation 
%he coupling parameter $L(\phi)$ is proportional to the exponential potential i.e., 
%$L(\phi)\propto V(\phi)\propto e^{\phi}$. Also, we have found that in
%the limit $\vert\beta\vert V(\phi)\ll \alpha$, the coupling function $L(\phi)$
%coincides with the expression for 
% $L(\phi)$ in the case when the integration constant $\beta=0$. 
 Also, we have obtained 
 that in order to satisfy that 
 the number of $e-$folds $N>0$,  we have imposed the condition $\alpha>\beta V$.

 To unify these results with the swampland criteria, 
 we have  found  a lower bound for the ansatz $\epsilon(N)=\epsilon_0=$ constant
  given by eq.(\ref{qqq}), together with the fact that the second swampland conjecture 
  $\epsilon_0>c^2/2\sim 1/2$. In 
  particular for the case in which $\Delta N=50$  and from the observational 
  data in which $\alpha=10^{10}$, we have obtained a range for the integration 
  constant $\beta$ when $\beta>0$ given by eq.(\ref{co1}) and for $\beta<0$, we 
  have otained the range described  by  eq.(\ref{co2}). Here from this unification, we have found that the range 
  for the parameter $\beta$ is very narrow, in order to satisfy 
  the observational data and swampland criteria. In particular  for the  
  specific case in which the integration constant 
$\beta=0$, we  have obtained that the effective potential   $V(N)\propto N$ and by 
combining  eqs.(\ref{gr8}) and (\ref{e1}), the consistency relation $r=r(n_S)$ becomes  
$r=4(1-n_S)$. Here we have found that the reconstruction of the  model does not work when the constant 
$\beta=0$.

Another ansatz on the slow roll parameter $\epsilon(N)$ that we have studied for large 
$N$, is given by eq.(\ref{AS2}), in which the parameter $\epsilon(N)\propto 1/N$. 
Interestingly in this situation, we have found that  the reconstruction for the effective potential $V(\phi)$ 
for $\beta>0$
coincides with the natural inflation. In this sense, we have obtained a pNGb 
potential given by eq.(\ref{V22}) in the reconstruction of this k-essence model. Here we have recognized the constants  $\beta^{-1/4}$
and $\sqrt{\alpha/(\beta \gamma)}$ as parameters associated to  mass scales from the particle physics 
models. For the case in which the integration constant $\beta<0$, we have found 
the effective potential given by 
 eq.(\ref{V23}) and this  potential corresponds to the hyperbolic inflation. 
 Here the hyperbolic potential has  a behavior of an exponential or power law 
 potential (quadratic potential) depending of  the limits applied to the scalar field.
 Additionally, we have obtained that for the 
specific case in which the integration constant $\beta=0$, the effective 
potential corresponds to a chaotic potential  in which $V(\phi)\propto\phi^2$.  
Also,  we have found that the reconstruction  for the coupling parameter $L(\phi)$ in the case in which 
 the integration constant $\beta>0$ is given by eq.(\ref{LL1})  and  for $\beta<0$  by 
 eq.(\ref{LL2}). 
 %In particular for the case in which $1\gg\vert \beta\vert V$, 
 %the coupling parameter $L(\phi)$ coincides with the special case in which $\beta=0$,  
% where $L(\phi)=$ constant.
 
As well, from the first  swampland conjecture we have obtain a lower limit for the parameter $\gamma$ 
given by eq.(\ref{ss}) and  by considering the second  conjecture we have found the  
limit $\gamma>N\,c^2/2$. In  particular for the  situation in which $\Delta N=50$, we 
have obtained numerically an upper bound for the constant $\beta$  given by $\beta<8.3\times 
10^8$ from both criteria. On the other hand, considering the observational data 
for the tensor to scalar ratio in which $r<0.07$, we have obtained a lower limit and then we have found 
that the  
range for 
the integration constant $\beta$ given by $1.5\times 10^8<\beta<8.3\times 10^8$. 
As before, we have found that the range for the parameter $\beta$ is very narrow 
from the unification of the observational data and swampland criteria.  In this respect, 
 these  narrow ranges on the parameter  $\beta$  found from this unification 
 suggest that the amplitude of primordial gravitational waves must have a lower limit 
 different from zero and similar to its upper bound. In this sense,  the results obtained 
 from the unification of the observational data and swampland criteria predict that the 
 amplitude of primordial gravitational waves can be detected in the future.

Thus, we have shown that it is possible to unify the theoretical foundations 
from the SC 
and the observational parameters corroborated by observations in 
the reconstruction of the early universe (inflation epoch).  In relation to  this point, 
we mention that the recontruction of inflation
only from the observational attractors does not guarantee that the SC are satisfied see 
e.g.\cite{Herrera:2018cgi,Herrera:2018mvo}.
In particular, it is possible an adaptation 
 to the first 
swampland criterion associated to  the range of the inflaton $\Delta\phi$, however 
this methodology of reconstruction only from the observational counterpart  does not ensure that the
second conjecture is satisfied.

Finally in this article, we have not addressed the reconstruction to
another  attractor point $n_S(N)$ or slow roll parameter 
$\epsilon(N)$. Also, we have not included the TCC in our analysis.  
 We hope
to return to these points in the near future.

\begin{acknowledgments}
 This work was supported by
Proyecto VRIEA-PUCV N$_{0}$ 039.449/2020.
\end{acknowledgments}

%\\\\\\\\\\\\\\\\\\\\\\\\\\\\\\\\\\\\\\\\\\\\\\\\\\\\\\\\\\\\\\\\\\\\\\\

\end{document}